\documentclass[aps,prb,twocolumn,floatfix,showpacs]{revtex4}
\usepackage[dvips]{graphicx}
\usepackage{array}
\usepackage{bm,amssymb,amsmath}
\usepackage{epsf}

\begin{document}

\title{Mott transition and dimerization in the one-dimensional SU$(n)$ Hubbard
model}

\author{K.~Buchta, \"O.~Legeza, E.~Szirmai, and J.~S{\'o}lyom}

\affiliation{Research Institute for Solid State Physics and Optics, H-1525
Budapest, P.\ O.\ Box 49, Hungary }

\date{\today}

\begin{abstract}
The one-dimensional SU$(n)$ Hubbard model is investigated numerically 
for $n=2,3,4$, and $5$ at half filling and $1/n$ filling using the density-matrix
renormalization-group (DMRG) method. The energy gaps and various quantum
information entropies are calculated. In the half-filled case, finite spin and
charge gaps are found for arbitrary positive $U$ if $n > 2$. Furthermore, it
is shown that the transition to the gapped phase at $U_{\rm c}=0$ is of
Kosterlitz-Thouless type and is accompanied by a bond dimerization both for
even and odd $n$. In the $1/n$-filled case, the transition has similar features 
as the metal-insulator transition in the half-filled SU(2) Hubbard model. The 
charge gap opens exponentially slowly for $U>U_{\rm c}=0$, the spin 
sector remains gapless, and the ground state is non-dimerized.
\end{abstract}


\maketitle

\section{Introduction}

Recently, the SU$(n)$-symmetric generalization of the standard SU(2) Hubbard
model\cite{Hubb1-4} has been intensively studied 
theoretically,\cite{marston,bozonos,bozonos_half,honer,assaad,szirmai01,szirmai02} since
this model may mimic strongly correlated electron systems where the orbital 
degrees of freedom of $d$ and $f$ electrons play important role.

Although the standard SU(2) Hubbard model is exactly solvable in one 
dimension\cite{BAM} by Bethe's ansatz and is well known to 
exhibit---at half filling---a Mott transition at $U_{\rm c}=0$, no 
such rigorous statement could be formulated for higher $n$ values. It is 
expected, however, that at generic fillings the one-dimensional SU($n$) 
Hubbard model behaves like an $n$-component Luttinger liquid while 
gaps may be generated at special fillings of the band, namely at half filling or 
$1/n$ filling.  

The one-dimensional half-filled SU($n$) Hubbard model has been
studied\cite{marston} in the large $n$ limit and it has been shown that the charge 
and spin modes---that are decoupled for $n=2$---become coupled and gap is 
generated in all of them. Moreover, it has been found that the system is 
spontaneously dimerized if $n$ is even. The role of um\-klapp processes in the 
half-filled model has been studied\cite{szirmai01} by the analytic multiplicative 
renormalization-group method in fermionic representation, too. It has been shown 
that in fact the um\-klapp processes  couple the spin and charge modes if $n>2$ 
and the spectrum is fully gapped for arbitrary values of the Coulomb repulsion. 
The question of dimerization has not been addressed in that work.

The $1/n$-filled case has been investigated\cite{bozonos} analytically using the
bosonized version of the model and numerically with quantum Monte Carlo simulation. 
It has been shown that at this special filling, i.e., when the number of particles is equal 
to the number of sites, the spin and charge degrees of freedom are decoupled and 
gap opens in the charge mode only. The spin modes remain gapless. Furthermore, it 
was inferred from the numerical calculations that for $n > 2$, unlike in the $n=2$ case, 
the charge gap opens at a finite positive $U_{\rm c}$. For smaller positive 
$U$ values the system shows metallic behavior. Since the contributions of the 
leading um\-klapp processes for commensurate fillings---except for half filling---are 
not logarithmically divergent, the special case of $1/n$ filling for $n>2$ could not 
be analyzed in the framework of the usual analytic renormalization-group procedure.

In this paper, we present a careful numerical analysis of the one-dimensional 
SU$(n)$-symmetric Hubbard model for $n=2,3,4$, and 5 using the density-matrix 
renormalization-group (DMRG) method.\cite{white} Besides the question where 
the Mott transition takes place in the $1/n$-filled model we will consider the problem 
of dimerization in the half-filled case, since Marston and Affleck\cite{marston} 
predicted dimerization for even $n$ only. This is done by calculating the one-site 
and two-site entropies\cite{zanardi,wu,gu,legeza_entropy} whose behavior may be a better 
indicator of where and how a quantum phase transition occurs than the study of 
opening of gaps, which is notoriously difficult for a Kosterlitz-Thouless transition.

The setup of the paper is as follows. The Hamiltonian is presented and the role of 
um\-klapp processes at commensurate fillings is discussed in Sec.\ II. A few questions 
concerning the numerical procedure used in the paper are presented in Sec.~III. 
In Sec.\ IV, the accuracy is tested on the SU(2) Hubbard model. The results 
obtained for half-filled and $1/n$-filled systems for $n > 2$ are given in 
Sec.~V and VI, respectively. The conclusions are summarized in Sec.~VII.

\section{The Hamiltonian and the role of um\-klapp processes}

The Hamiltonian of the SU$(n)$ Hubbard model is written in the form
\begin{equation} \begin{split}
 \label{eq:ham}
  {\mathcal H} & =   - t\sum_{i=1}^N\sum_{\sigma=1}^n (c_{i,\sigma}^\dagger
     c_{i+1,\sigma}^{\phantom \dagger} + c_{i+1,\sigma}^\dagger
    c_{i,\sigma}^{\phantom\dagger}) \\
       & \phantom{=}  +  \frac{U}{2}\sum_{i=1}^N
  \sum_{\substack{\sigma,\sigma'=1 \\ \sigma\neq\sigma'}}^n n_{i, \sigma}n_{i,
     \sigma'} ,
\end{split}   \end{equation}
where $N$ is the number of sites in the chain, $c_{i,\sigma}^\dagger$
($c_{i,\sigma}^{\phantom\dagger}$) creates (annihilates) an electron at site
$i$ with spin $\sigma$, the spin index is allowed to take $n$ different values, 
$n_{i,\sigma}$ is the particle-number operator, $t$ is the hopping integral between 
nearest neighbor sites, and $U$ is the strength of the on-site Coulomb repulsion. 
In what follows $t$ will be taken as the unit of energy. The dimensionless $U$ and 
the dimensionless gaps used in the paper are obtained by dividing their physical values by $t$. 

The term $\sigma = \sigma'$ could have been kept in the last summation, as it is 
usually done in the literature, in order to display clearly the SU($n$) symmetry.
We prefer this form where the Coulomb repulsion acts between particles of 
different spin index only. Since it is forbidden to have two particles of the same 
spin on the same site, the two expressions differ in a shift in the energy only. 

At generic fillings, this model is equivalent to an $n$-component Luttinger 
liquid. It has one symmetric (charge) and $n-1$ antisymmetric (spin) gapless 
bosonic modes. At special fillings, where um\-klapp processes may become 
relevant, we will be interested especially in the half-filled and $1/n$-filled
cases, gap may be opened in the spectrum of charge or spin excitations. 
In this section we will address the question how the charge and spin degrees 
of freedom are coupled by um\-klapp processes.

Let us consider $p/q$-filled systems where $p$ and $q$ are relative prime
integers. Due to particle-hole symmetry it suffices to consider the case $p < q/2$. 
At such a filling the Fermi momentum is $k_\textrm{F}=\pi p/q$ (the lattice
constant has been taken to be unity). Although the bare Hamiltonian contains 
two-particle scatterings only, $l$-particle scattering processes may appear
as higher-order perturbations.\cite{giamarchi} If $l$ particles are scattered from one
of the Fermi points to the opposite one, the total change in the momentum is
\begin{equation}
   \Delta k= \pm 2k_\mathrm{F} l = \pm 2 \pi l p/q \,.
\end{equation}
Such um\-klapp processes are allowed by momentum conservation if $\Delta k$ 
is an integer multiple of $2\pi$, i.e., $l$ has to be a multiple of $q$. The leading, 
lowest-order um\-klapp processes correspond to $l=q$. Since the interaction is 
assumed to be local in real space, in the dominant um\-klapp processes all scattered 
particles have to have different spin indices, i.e., we will assume that $l \leq n$. 

Since the most significant contribution to low-lying excitations comes from
fermion states close to the Fermi points, a linearized spectrum is assumed. The 
relevance or irrelevance of um\-klapp processes and the coupling 
between different modes can then be conveniently analyzed by transforming the
Hamiltonian into bosonic form.\cite{bozonizacio} Using the phase
field $\phi_{\sigma}$ for particles with spin $\sigma$, which is the sum of the
phases of the right- and left-moving fermions, the $l$-particle um\-klapp processes 
can be represented by an effective term in the Hamiltonian which is proportional to
\begin{equation}
\label{eq:umkl}
\int \textrm{d}x \sum_{\{\sigma_i\}'}
\cos\big[2\left(\phi_{\sigma_1}(x)+\ldots +\phi_{\sigma_l}(x)\right)\big] ,
\end{equation}
where $\{ \sigma_i\}'$ indicates that all spin indices are assumed to be different.

In order to investigate the role of these processes let us introduce---in 
the usual way---the symmetric and antisymmetric combinations of the bosonic 
fields for which the term charge and spin modes, respectively, will be used:
\begin{eqnarray}
\label{eq:modes}
\phi_\mathrm{c}(x) & = & \frac{1}{\sqrt{n}}\sum_{\sigma=1}^n\phi_\sigma(x), \\
\phi_{m\mathrm{s}}(x) & = &\frac{1}{\sqrt{m(m+1)}}\left( \sum_{\sigma=1}^m
\phi_\sigma(x) - m\phi_{m+1}(x) \right),    \nonumber
\end{eqnarray}
with $m=1,\ldots ,n-1$.

When the band is $p/n$ filled where $p$ is relative prime to $n$, the leading um\-klapp 
term corresponds to scattering of $n$ particles from one of the Fermi points to the
opposite one. In this case, the effective term in the Hamiltonian is proportional to
\begin{equation}
\int \textrm{d} x \cos \big[2\sqrt{n}\phi_{\rm c}(x)\big] .
\end{equation}
As can be seen, this term involves the charge sector only. On the other hand, for 
a $p/q$-filled band with $q < n$, several terms may appear in \eqref{eq:umkl} 
as leading ($l=q$) um\-klapp processes. When written in terms of the bosonic 
fields, both charge and spin modes appear in them. For example in the SU(3) 
model, where in addition to the charge mode there are two spin modes, in the 
half-filled case, the contribution of the two-particle um\-klapp processes is 
proportional to
\begin{eqnarray}    
      \int \textrm{d} x \! &\! \bigg( \! & \! \cos\bigg[\sqrt{\frac{2}{3}}\left( 2 \sqrt{2}
    \phi_{\textrm{c}}(x)+ 2\phi_{\textrm{2s}}(x)\right)\bigg]   \\ 
    & \!+ \!& \! \cos\bigg[ \sqrt{\frac{2}{3}}\left( 2 \sqrt{2}\phi_{\textrm{c}}(x) + 
       \sqrt{3}\phi_{\textrm{1s}}(x) - \phi_\textrm{2s}(x) \right) \bigg]  \nonumber \\ 
     &\! + \!& \!\cos\bigg[ \sqrt{\frac{2}{3}}\left( 2 \sqrt{2}\phi_{\textrm{c}}(x)-
     \sqrt{3}\phi_{\textrm{1s}}(x) - \phi_\textrm{2s}(x) \right) \bigg]\bigg). \nonumber
\end{eqnarray} 
These terms are relevant and the corresponding soliton excitations couple 
the charge and spin modes. Therefore, gap is generated not only in the charge 
sector but in the spin sector as well, if $q < n$, in agreement with Refs.\ 
\onlinecite{marston} and \onlinecite{szirmai01}.

\section{Numerical procedure}

\subsection{Numerical accuracy}

The numerical calculations presented in this paper have been performed on finite chains 
with open or periodic boundary condition (OBC or PBC, respectively) using the 
DMRG technique,\cite{white} and the dynamic block-state selection (DBSS) 
approach.\cite{legeza02,legeza03} All data shown in the figures were obtained with 
OBC unless stated otherwise.  We have set the threshold value of the quantum 
information loss $\chi$ to $10^{-5}$ and the minimum number of block states 
$M_{\rm min}$ to $256$. All eigenstates have been targeted independently 
using four DMRG sweeps until the entropy sum rule has been satisfied.  The 
accuracy of the Davidson diagonalization routine has been set to $10^{-7}$ 
and the largest dimension of the superblock Hamiltonian was around three millions.

In the DBSS procedure the DMRG parameters are set dynamically. The maximum 
number of block states ($M_{\rm max}$) that our program could handle was $2500$, 
$1500$, $800$, and $256$ for $n=2,3,4$, and $5$, respectively. This determines the 
maximal chain  length that could be treated reliably with the accuracy prescribed in terms 
of $\chi$. For small $U$ $(U <1)$, where the coherence length is large, the block 
entropy grows very rapidly with increasing block size and the upper cutoff on the number 
of block states is reached at $N\simeq 90$ for the SU(3) model and at 
$N\simeq 30$ for the SU(4) model. For these $U$ values and for such chain lengths the 
truncation error was of the order of $10^{-6}$ for a few DMRG iteration steps and the 
absolute error of our calculation is in the range of $10^{-4}$. Calculations on longer 
systems would give less reliable results and this imposes a serious limitation on the accuracy 
of the results obtained by finite-size extrapolation.
 
\subsection{Detecting and locating phase transitions}

The most common procedure to locate quantum phase transitions numerically is
to calculate energy gaps. If the SU$(n)$ symmetry is not broken in the ground state
and the band is $p/q$ filled, in a system with $N$ lattice sites, the number of particles 
with spin index $\sigma$ is $N_{\sigma}= Np/q$. The ground-state energy is 
denoted by $E_0(N)$. The spin and charge gaps corresponding to the increase of 
energy when changing the spin of a particle or changing the number of particles were 
calculated according to the formulae
\begin{equation}   \begin{split}
     \Delta_{\text{s}}(N) & =  E_{+1,-1}(N) - E_0(N) \,,    \\ 
     \Delta_{\text{c}}(N) & =  E_{+1}(N)  + E_{-1}(N) - 2 E_0(N) \,,
\end{split}  \end{equation}
where $E_{+1}(N)$ is the lowest energy eigenvalue of the Hamiltonian when 
$N_{\sigma}$ is increased by one for a given spin, $E_{-1}(N)$ is the lowest 
energy when $N_{\sigma}$ is decreased by one for a given spin, and 
$E_{+1,-1}(N)$ is the lowest energy when the number of particles with spin 
$\sigma$ is increased by one while the number of particles with
a different $\sigma'$ is decreased by one. 

Since---as will be seen---it is difficult to study numerically the closing of energy gaps 
for small $U$ values, an alternative approach to study quantum phase transitions
has been suggested by several groups.\cite{zanardi,wu,gu,legeza_entropy} It uses 
the anomalies appearing in the generalized $l$-site entropy functions. These functions
are easily accessible in DMRG and require to target the ground-state wavefunction only. 
Moreover, they are expected to have better finite-size scaling properties than the energy gaps.

The von Neumann entropy of a single site can be determined from 
$s_i = -{\rm Tr}  \rho_i \ln \rho_i$, where the reduced density matrix $\rho_i$ of 
site $i$ is obtained from the wavefunction of the total system by tracing out all 
configurations of all other sites. In a similar manner generalized $l$-site entropies can 
be calculated which are often better indicators of quantum phase transitions than the 
one-site entropy. In our DMRG approach, the one- ($s_i$) and two-site ($s_{i, i+1}$) 
entropies at the center of the chain, for $i=N/2$ or $i=N/2+1$ and the block entropy 
of the left half of the system (corresponding to $l=N/2$) are calculated at the end 
of each DMRG sweeps.

It turned out that for the SU($n$) Hubbard model the same single-site entropy is obtained 
when it is calculated on neighboring sites in the center of the chain. The sites are equivalent.
The two-site entropy $s_{i,i+1}$ is, however, different when it is considered at $i=N/2$ 
or $i=N/2+1$. An indication of the existence of bond dimer order, the breaking of 
translational symmetry can be obtained---as an alternative to the usual dimer order 
parameter---from the difference of two-site entropies,
\begin{equation}
     D_s(N) = s_{N/2,N/2+1}-s_{N/2+1,N/2+2} \,.
\label{eq:d_s}
\end{equation}
As it has been shown in Ref.\ \onlinecite{buchta01} and will be discussed below, 
usually the dimer entropy difference converges faster than the energy gap, and it may be 
more convenient to analyze this quantity. In the next sections, we will use $D_s$ to 
study the phase diagram as a function of $U$.

As has been pointed out recently by two of the authors,\cite{legeza_incomm} further 
information about possible nonuniform phases can be obtained from the study of
the length dependence of the von Neumann entropy of a block of $l$ sites in a finite
chain, $s(l), l=0,\cdots,N$, and its Fourier spectrum, 
\begin{equation}
      \tilde{s}(q) = \frac{1}{N}\sum_{l=0}^N e^{- i q l}s(l)\,.
\end{equation}
A finite peak at a nonzero wave vector indicates a corresponding modulation of the 
state. E. G., in a dimerized (trimerized) state the $q=\pi$ ($q=2\pi/3$) Fourier 
component is nonvanishing. Furthermore, in a gapless model the central charge 
can be derived\cite{holzhey,cardy,affleck,laflorence} from the initial slope
of the length dependence of $s(l)$. This can help to distinguish better gapped and 
gapless regimes.  
 

\subsection{Finite-size scaling}

The large-$N$ limit of the energy gaps and entropies can be obtained if appropriate
scaling functions are used. In a critical, gapless model, in leading order, the gap 
$\Delta(N)$ is expected to scale to zero as $1/N$. In a non-critical model the 
scaling depends on the boundary condition. The leading correction is exponential,
if PBC is used:  
\begin{equation}
    \Delta(N) = \Delta + c{1\over N^{1/2}}\exp(-N/\xi),
    \label{eq:scale_pbc}
\end{equation}
while for OBC the corrections are algebraic, and $\Delta(N)$ is expected
to vary as
\begin{equation}
    \Delta(N) = \Delta + a/{N^2} + {\cal O}(N^{-4}).
    \label{eq:scale_obc}
\end{equation}
Therefore, the following fitting ansatz was used to evaluate the results obtained
with OBC,
\begin{equation}
    \Delta(N) = \Delta + a/N + b/{N^2},
    \label{eq:scale_N}
\end{equation}
with $\Delta$, $a$, and $b$ as free parameters. 

Local quantities, however, are expected to have a better scaling property even
for OBC.  For non-critical models the end effects decay exponentially with a finite 
correlation length, and the leading correction to the local quantities $s_{N/2}(N)$, 
$s_{N/2, N/2+1}(N)$, and $D_s(N)$ is expected to have the form
\begin{equation}
    A(N) = A + d N^{-\beta} \exp(-N/2\xi),
    \label{eq:scale_D}
\end{equation}
where $A$ is any of the local quantities listed above. This form is qualitatively 
similar to (\ref{eq:scale_pbc}), except that $N/2$---the distance of the middle 
of the chain from the boundary---appears in the exponential and the exponent 
of the algebraic prefactor is \emph{a priori} unknown.

\section{The SU(2) model as a reference system}

Since numerical accuracy and proper finite-size scaling are crucial in the present work,
we have tested our approach first on the half-filled SU(2) Hubbard model, where exact
results are available. 

\begin{figure}[htb]
\includegraphics[scale=0.5]{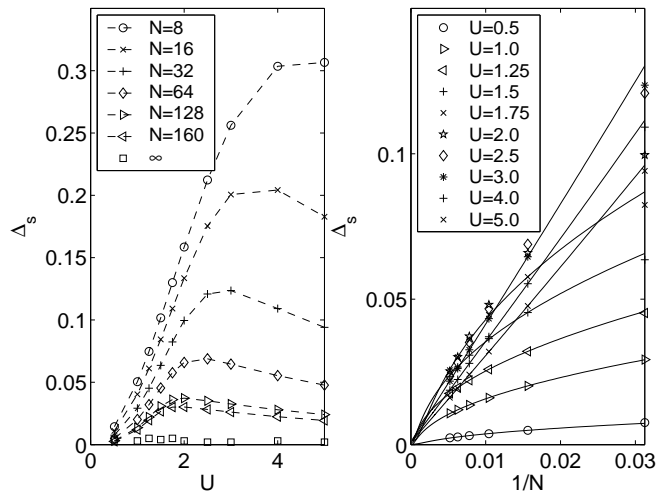}
\caption{$U$ dependence and finite-size scaling of the spin gap for  the half-filled 
SU(2) Hubbard model. The dashed lines are guide to the eye, the solid lines show 
the result of our fit.}
\label{fig:su2_spin_half}
\end{figure}

\begin{figure}[htb]
\includegraphics[scale=0.5]{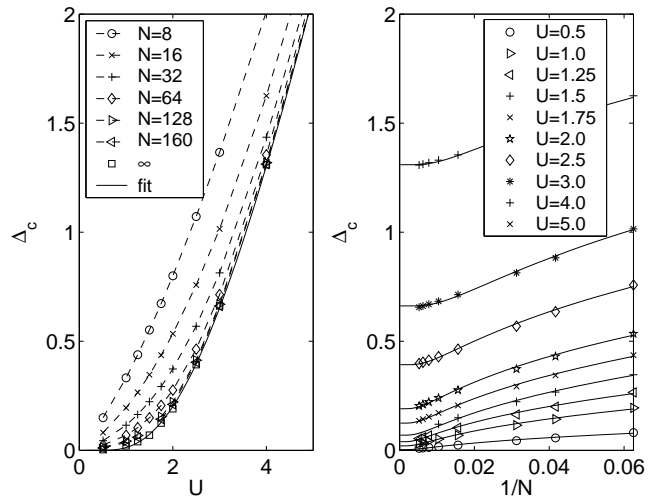}
\caption{$U$ dependence and finite-size scaling of the charge gap for the half-filled 
SU(2) Hubbard model. The dashed lines are guide to the eye, the solid lines show 
the result of our fit.}
\label{fig:su2_charge_half}
\end{figure}

The energy of all eigenstates that have been determined using DMRG with PBC 
agreed up to 5 digits with the numerical solution of the Bethe-ansatz equations.  
The $U$ dependence of the spin and charge gaps and the finite-size extrapolation
are shown in Figs.~\ref{fig:su2_spin_half} and \ref{fig:su2_charge_half},
respectively.

Although the spin gap has a maximum as a function of $U$, this is a finite-size
effect. Using (\ref{eq:scale_pbc}) for the extrapolation to the thermodynamic limit,
the spin gap scales to $\Delta_{\rm s} = 6(1)\times 10^{-4}$ for small and large $U$ 
while it is of the order of $10^{-3}$ in the intermediate region. Here and in what 
follows the digits in parentheses are the one-standard-deviation uncertainty in the last 
digits of the given value. The root-mean-square error (the norm of residuals) of the fit 
denoted by $\kappa$ was $10^{-6}$.  

The charge gap decreases monotonically with decreasing $U$, and
the extrapolated values obtained using (\ref{eq:scale_pbc}) are finite, 
opening exponentially slowly, 
\begin{equation}
\Delta(U) = a \exp[-c(U-U_{\rm c})^{-\sigma}] ,
\label{eq:kt}
\end{equation}
which is characteristic to a Kosterlitz-Thouless transition. The non-universal constants 
$a$ and $c$ and the exponent $\sigma$ were estimated together with $U_{\rm c}$ 
by a least square fit. The best fit with error $\kappa = 5\times 10^{-4}$ could be 
achieved with $a=100.513$, $c=8.931$, $\sigma = 0.517$, and $U_{\rm c}=0.024$.
If $\sigma$ is fixed to $\sigma=0.5$ the parameters $a$ and $c$ change only slightly,
and the best fit with error $\kappa = 8\times 10^{-4}$ gives $U_{\rm c}= 0.075$. 
This can be taken as an indicator of the accuracy since the 
exact result is $U_{\rm c}=0$.

\begin{figure}[htb]
\includegraphics[scale=0.35]{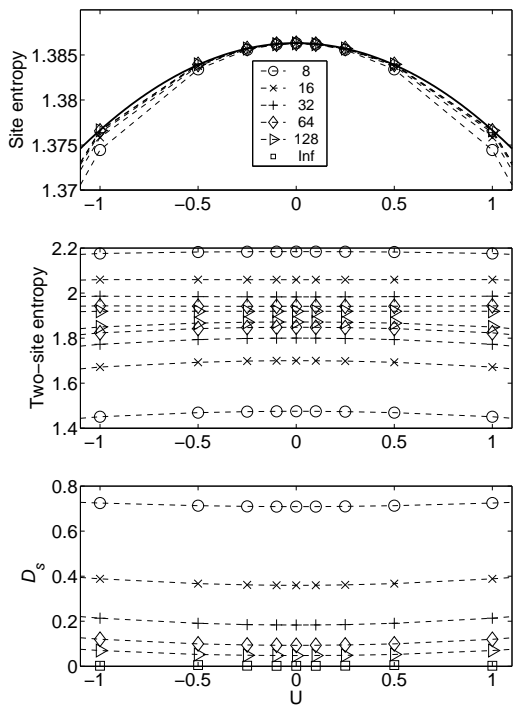}
\caption{One-site and two-site entropy and dimerization of the two-site entropy 
in the middle of the chain as a function of $U$ for the half-filled SU(2) Hubbard model.
The two sets of curves for the two-site entropy correspond to $i=N/2$ and
$i=N/2+1$. The dashed lines are guide to the eye and the solid line is our parabolic fit.} 
\label{fig:entropy_su2}
\end{figure}

The one-site and two-site entropies measured in the middle of the chain as well 
as the dimerization in the two-site entropy obtained for chains with up to $N=128$ 
sites as a function of $U$ are plotted in Fig.~\ref{fig:entropy_su2}.  
It is worth mentioning that the block entropy measured for the symmetric 
superblock configuration shows similar behavior as the two-site entropy.
 
The one-site, two-site and block entropies take their maximal value at $U=0$, corresponding
to the equipartition of local states,\cite{gu,larson} and no anomaly can be seen 
for $U>0$ in agreement with the known analytic result.\cite{gu}  The one-site entropy 
could be fitted well with a parabola 
\begin{equation}
          s_{N/2} = s_0 - A U^2\,,
\end{equation}
yielding $s_0 = 1.3863$, $A = 0.009293$, with error $\kappa = 5 \times 10^{-9}$. 
This parabolic fit is also shown in Fig. \ref{fig:entropy_su2}. The constant $s_0$
and the coefficient $A$ are in very good agreement with the exact result, 
$s_0 = \ln 4 $ and $A= [7\zeta(3)/2\pi^3]^2/2 = 0.0092057$.

Although the two-site-entropy is markedly different on neighboring bonds in finite
systems, the dimerization of the two-site entropy defined by (\ref{eq:d_s}) scales 
to $D_s = 5(1) \times 10^{-4}$ for all $U$, which should be considered as 
zero, in agreement with the known result that the SU(2) Hubbard model is not
dimerized.

From these calculations we conclude that a value less than about $5\times 10^{-4}$
either for the gap or the dimerization of the two-site entropy should be taken as zero. 

\section{Half-filled SU$(n)$ models}

The renormalization-group calculations\cite{marston,szirmai01} have shown that at
half filling the charge mode is always gapped. When $n > 2$, this mode is coupled to 
the spin sector and this generates gaps in the spin modes, as well. Moreover, Marston 
and Affleck\cite{marston} have pointed out that for even $n$ the ground state is a 
charge-density wave state where the charge density is centered on the bonds, i.e., 
the system is spontaneously dimerized.

Since the arguments leading to this result cannot be straightforwardly extended to 
odd $n$ we have done calculations both for even and odd $n$ to compare the 
behavior of the SU($4$) Hubbard model to that of the SU($3$) and SU($5$) models. 

\subsection{Models with even $n$}

As has been shown, in the SU(2) model already, where relatively long chains 
could be studied, the energy gaps could not be determined for small $U$ values with 
better accuracy than $5\times 10^{-4}$. Since the upper cutoff on the number of 
DMRG block states is reached for relatively small system sizes, $N=32$ already
in the SU(4) model, determination of the $N \rightarrow \infty$ limit of the energy 
spectrum with the same accuracy is not possible with our computational facilities. 
We have calculated the energy gaps for a few large $U$ values, where longer chains 
up to $N=64$ could be treated. Although the extrapolated gaps are somewhat larger 
than those reported in Ref. \onlinecite{bozonos_half}, e.g., for $U=4$ we have found
$\Delta_{\rm c} = 0.69(4)$ and $\Delta_{\rm s} = 0.26(2)$, the results demonstrate 
clearly that in fact both gaps are finite. 

\begin{figure}[htb]
\includegraphics[scale=0.35]{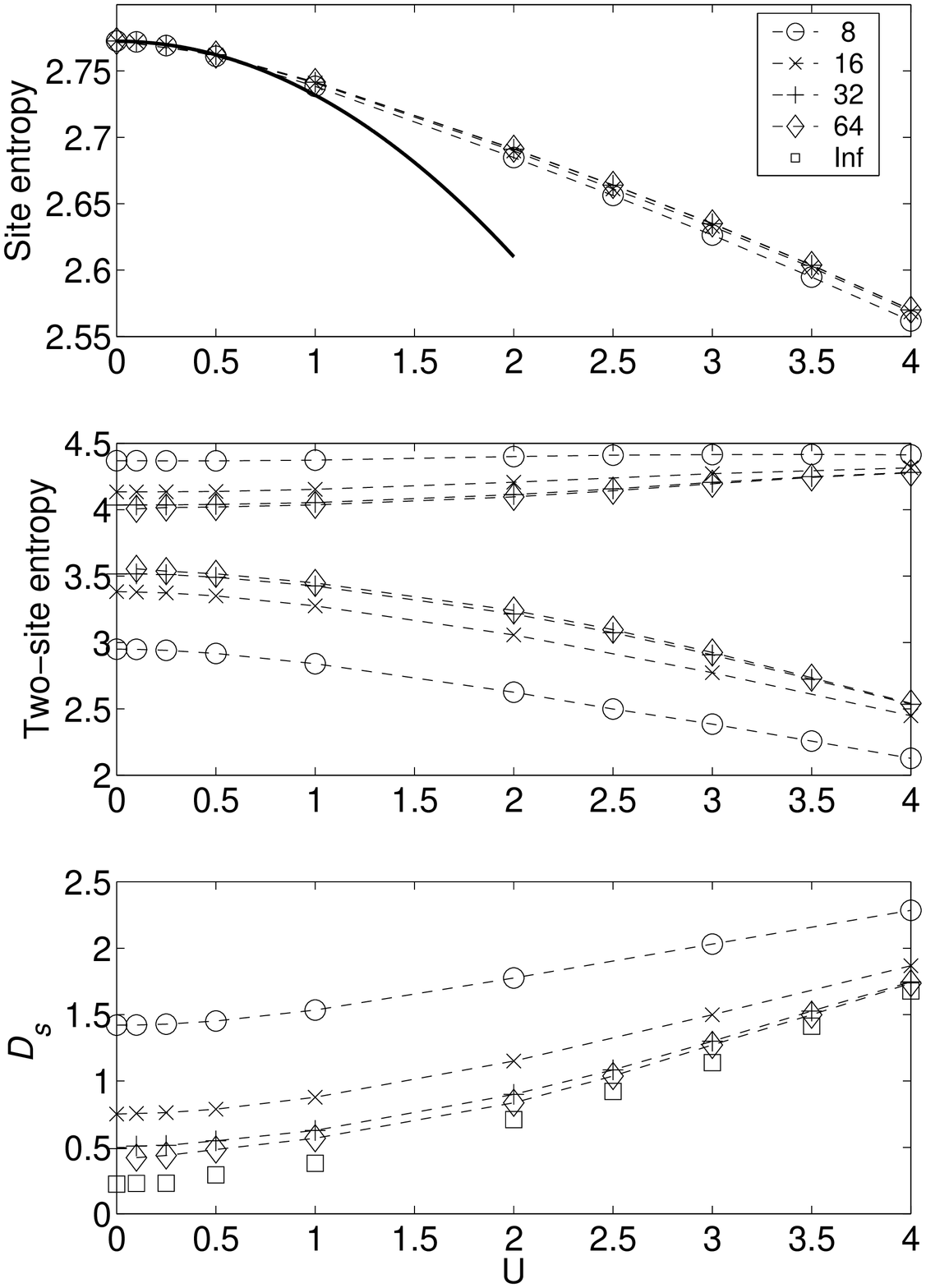}
\caption{Entropy functions plotted as in Fig.\ \ref{fig:entropy_su2}, but for 
the half-filled SU(4) Hubbard model.}
\label{fig:entropy_su4_half}
\end{figure}

Since our primary aim for the half-filled case was to study dimerization, we have 
analyzed the ground state entropy functions. The one-, and two-site entropies and
the dimerization of the two-site entropy are shown in Fig.~\ref{fig:entropy_su4_half} 
for different chain lengths. While the one-site entropy shows no sign of dimerization, 
bond dimerization is apparent in the two-site entropy.  
Unfortunately the limitation for small $U$ values 
discussed above applies also in the case when the extrapolated values of $D_{s}$ 
are calculated. Although there is no doubt that $D_s$ is finite for not too small $U$, 
and in the large-$N$ limit, the dimerization of the two-site entropy seems to grow 
exponentially slowly as a function of $U$, resembling a Kosterlitz-Thouless transition,
no reliable finite-size scaling analysis could be done to determine $U_{\rm c}$ 
where $D_s$ and the energy gaps become finite. The reason is that if a smaller 
$\chi$ is required in the calculation for the longest chain with $N=64$ sites, smaller 
$D_s$ is obtained, thus our results shown for $N=64$ overestimate $D_s$.

We try to infer $U_{\rm c}$ from the behavior of the one-site entropy. As seen, it is 
a continuous function of $U$ with a maximum at $U=0$ and without any anomaly for 
$U>0$ 
For small $U$ values it can be fitted well (with error $\kappa=1 \times 10^{-7}$) 
by a parabola with $s_0 = 2.7723$ (the exact value is $s_0 = \ln 16$) and 
$A = 0.0404$. 

This analytic behavior and the results obtained for the dimerization
are in complete agreement with the prediction by Marston and Affleck.\cite{marston}

\subsection{Models with odd $n$}

\begin{figure}[htb]
\includegraphics[scale=0.5]{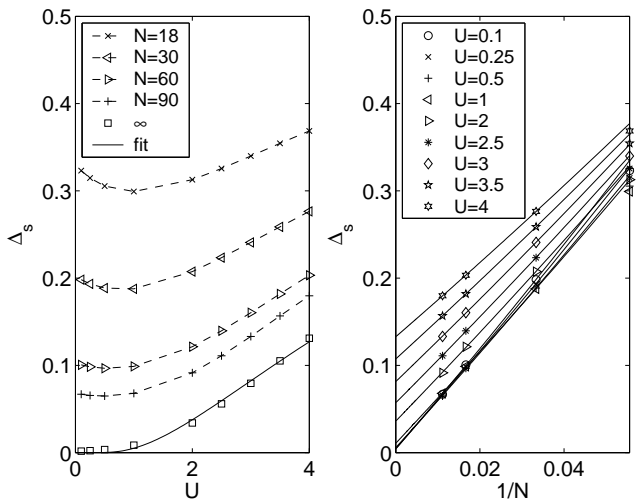}
\caption{The spin gap plotted as in Fig. \ref{fig:su2_spin_half}, but for the 
half-filled SU(3) Hubbard model.}
\label{fig:su3_spin_half}
\end{figure}

\begin{figure}[htb]
\includegraphics[scale=0.5]{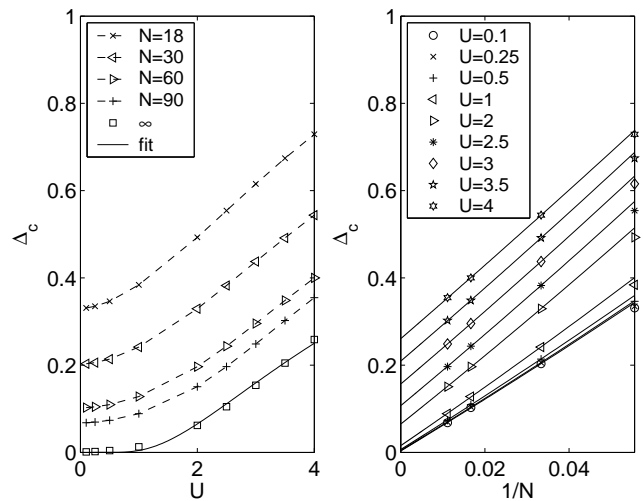}
\caption{The charge gap plotted as in Fig. \ref{fig:su2_charge_half}, but for the 
half-filled SU(3) Hubbard model.}
\label{fig:su3_charge_half}
\end{figure}

We will now compare the results described above with the behavior of
the SU($n$) model for odd $n$. The $U$ dependence of the spin gap obtained by
DMRG for the SU(3) half-filled Hubbard model and the finite-size extrapolation
are shown in Fig.~\ref{fig:su3_spin_half}. The corresponding plots for the charge gaps 
are shown in Fig.~\ref{fig:su3_charge_half}.

The extrapolated values obtained using (\ref{eq:scale_N}) are summarized in 
Table~\ref{table:su3_half}. They were fitted with a four-parameter curve (\ref{eq:kt}). 
The least-square fit (also shown in Fig.\ \ref{fig:su3_spin_half}) for the spin gap
with error $\kappa = 8\times 10^{-7}$ gives $a = 1.089$, $c= 4.9256$, 
$\sigma = 0.607$, and $U_{\rm c} = 0.054$. When $\sigma$ is fixed to 
$\sigma=0.5$ we obtain  $U_{\rm c} = 0.098$ with error $\kappa= 2\times 10^{-6}$. 

For the charge gap the same procedure gives $a=1.0855$, $c= 5.0492$,
$\sigma = 0.9060$ and $U_{\rm c}= 0.0913$ with error $\kappa = 5\times 10^{-7}$
(this curve is also shown in Fig.\ \ref{fig:su3_charge_half}) or $U_{\rm c}= 0.099$ 
with error $\kappa = 2\times 10^{-5}$ if $\sigma=0.5$.

\begin{table}[htb]
\begin{tabular}{@{\hspace{1mm}}lc@{\hspace{5mm}}c@{\hspace{4mm}}c@{\hspace{1mm}}}
\hline \hline
\multicolumn{1}{c}{$U$} &  & $\Delta_{\rm{c}}$ & $\Delta_{\rm{s}}$ \\
\hline
0.1  &  & 0.0010(5)  &       0.0017(4)  \\
0.25 &  & 0.0016(4)  &       0.0021(4)  \\
0.5  &  & 0.0040(4)  &       0.0035(4)  \\
1    &  & 0.0128(3)  &       0.0086(3)  \\
2    &  & 0.0618(2)  &       0.0339(3)  \\
2.5  &  & 0.1047(1)  &       0.0559(2)  \\
3    &  & 0.1537(2)  &       0.0794(2)  \\
3.5  &  & 0.2047(1)  &       0.1057(1)  \\
4    &  & 0.2586(1)  &       0.1318(1)  \\
\hline \hline
\end{tabular}
\caption{Extrapolated values of the spin and charge gaps in the thermodynamic
limit for the half-filled SU(3) Hubbard model.}
\label{table:su3_half}
\end{table}

\begin{figure}[htb]
\includegraphics[scale=0.35]{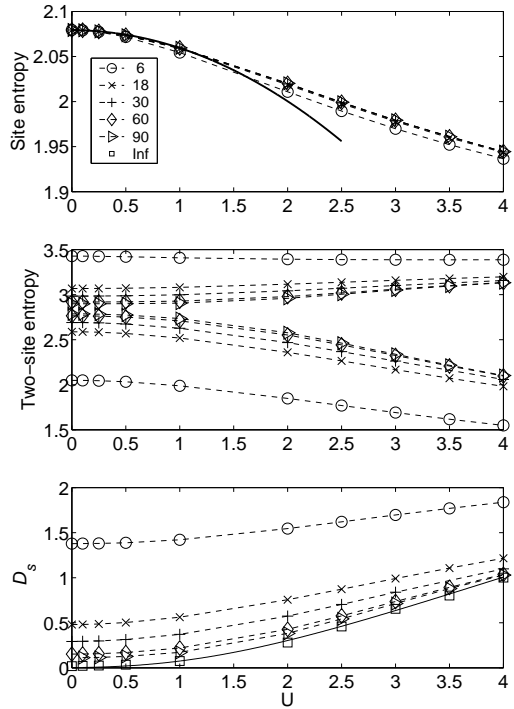}
\caption{Entropy functions plotted as in Fig.\ \ref{fig:entropy_su2}, but 
for the half-filled SU(3) Hubbard model.}
\label{fig:entropy_su3_half}
\end{figure}

The one- and two-site entropies and the dimerization appearing in the latter one
are shown in Fig.~\ref{fig:entropy_su3_half} for system sizes up to $N=90$. 
The one-site entropy possesses a maximum at $U=0$, and none of the entropy 
functions show any sign of anomaly for $U>0$. The one-site entropy can be fitted 
with a parabola giving $s_0=2.0791$ (the exact value is $s_0 =\ln 8$) and $A=0.0197$ 
with $\kappa=5\times 10^{-7}$. Bond dimerization is signaled again by the two-site 
entropy, since $D_s$ 
remain finite in the large-$N$ limit for all finite $U$ values. 
Eq.~(\ref{eq:kt}) with fixed $\sigma = 0.5 $ gives $a=18.324$, $c=5.827$, and 
$U_{\rm c}=0.013$ with error $\kappa = 5\times 10^{-7}$.

\begin{figure}[htb]
\includegraphics[scale=0.5]{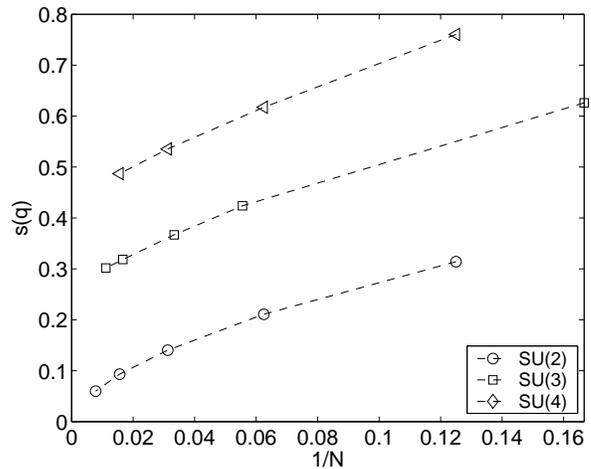}
\caption{Finite-size scaling of the $q=\pi$ Fourier component of the 
half-filled SU($n$) Hubbard model for $n =2,3,4$ at $U=4$.} 
\label{fig:sun_fft}
\end{figure}

This finding, the dimerization of the half-filled SU$(3)$ model has been
corroborated by the study of the length dependence of the block entropy. We have 
determined the von Neumann entropy of blocks of length $l$ for various chain
lengths and calculated the Fourier components. It was found that an oscillatory
component is superimposed on the smoothly increasing $s(l)$ as $l$ varies from
$l=0$ till $l=N/2$ (due to its definition $s(l)$ decreases for $l > N/2$ to vanish at
$l=N$) and a finite Fourier component at $q=\pi$ is obtained in the $N \rightarrow \infty$
limit. This is shown in Fig. \ref{fig:sun_fft}. Comparison to the results obtained for 
the half-filled SU(2) and SU(4) models, also shown in the figure, it is clearly seen
that the SU(3) model behaves like the SU(4) model, both are dimerized, while the 
SU(2) model is not.

Due to the limitations imposed by our computing facility we could calculate the 
entropy functions for the SU(5) Hubbard model for chains up to $N=20$ lattice 
sites only. Therefore these results show tendencies only. It was found that in 
the half-filled case the one-site and two-site entropies have a smooth maximum at 
$U=0$ and the dimerization of the two-site entropy as a function of $U$ 
bf as well as the oscillation of the block entropy show  
similar behavior as in the half-filled SU(3) and SU(4) models: it is finite, the system
is dimerized.

From all these results we conclude that the half-filled SU($n$) Hubbard models
with odd $n$ behave in the same way as predicted for even $n$. 

\section{$1/n$-filled SU$(n)$ models}

The other interesting situation which in a certain sense is the analogon of the
half-filled SU(2) model is the $1/n$-filled band, where the number of electrons 
is equal to the number of lattice sites. In this case um\-klapp processes in which $n$ 
particles with different spin indices are scattered across from $k_{\rm F}$ to 
$-k_{\rm F}$ (or vice versa) may become relevant. An earlier study\cite{bozonos} 
of the model using both analytic and numerical quantum Monte Carlo approach 
suggested that the spin gap vanishes for all $U$, while the charge gap opens 
exponentially slowly for $U>U_{\rm c}$ but with finite $U_{\rm c}$. Since these numerical 
calculations have been performed for relatively short chains up to $N=30$ sites,
in this section we investigate in detail the same problem for the SU(3) model on 
longer chains using a different and hopefully more accurate procedure, and check 
the results for SU(4) and SU(5) models, too.
  
\subsection{SU(3) model at one-third filling}

\begin{figure}[htb]
\includegraphics[scale=0.5]{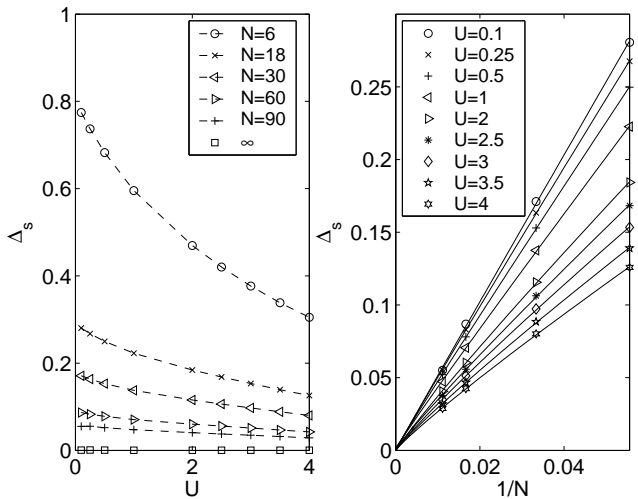}
\caption{Spin gap plotted as in Fig. \ref{fig:su2_spin_half}, but for the 
one-third-filled SU(3) Hubbard model.}
\label{fig:su3_spin_third}
\end{figure}

\begin{figure}[htb]
\includegraphics[scale=0.5]{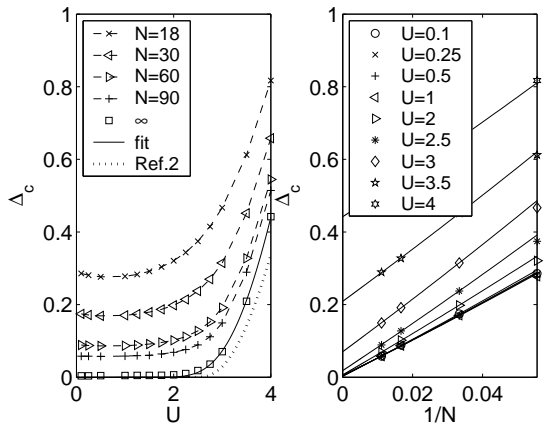}
\caption{Charge gap plotted as in Fig. \ref{fig:su2_charge_half}, but for 
the one-third-filled SU(3) Hubbard model.} 
\label{fig:su3_charge_third}
\end{figure}

The $U$ dependence of the spin gap obtained by DMRG for the SU(3) one-third-filled 
Hubbard model and the finite-size extrapolation are shown in Fig.~\ref{fig:su3_spin_third}.  
The corresponding plots for the charge gaps are shown in Fig.~\ref{fig:su3_charge_third}.

As it is seen, for the chain lengths available the spin gap is a decreasing function of $U$ 
and it scales to zero slower than $1/N$. An upper bound $\Delta_{\rm s} = 0.003(2)$ is 
obtained for all $U$ if (\ref{eq:scale_N}) is used for the extrapolation. A better fit 
can be achieved by the ansatz 
\begin{equation}
    \Delta(N) = \Delta + a/N^b, 
    \label{eq:scale_N_alpha}
\end{equation}
with the exponent $b$ as a free parameter. This fit gives $\Delta_{\rm s} = 0.0006(3)$ 
with $b$ varying between $0.9$ and $0.98$. This difference is an
indicator of the limits of our calculations. As mentioned earlier a gap $5\times10^{-4}$ 
should be taken to be zero. Thus we conclude that the spin gap vanishes in the 
large-$N$ limit for all positive $U$. 

In contrast to this, the charge gap is a monotonically increasing function of
$U$ for all finite system sizes. It scales to finite values not only if $U$ is large 
enough but for small $U$ values, as well. Our finite-size scaling analysis predicts 
small but finite charge gap. The extrapolated values are given in 
Table~\ref{table:su3_third}. These values are larger than those given in Ref.\ 
[\onlinecite{bozonos}] where the longest chain used in the finite size scaling analysis 
had $N=30$ sites. The reason for the discrepancy is that the $1/N^2$ corrections 
in Eq.~(\ref{eq:scale_obc}) can be seen for longer systems only. 
The extrapolated values were fitted with the four-parameter function 
(\ref{eq:kt}). The best estimates for the parameters are: $a = 8.68$, $c= 28.12$, 
$\sigma=1.62(3)$, and $U_{\rm c} = 0.03(3)$  (this curve is also shown in 
Fig.\ \ref{fig:su3_charge_third}).  The error of the fit is $\kappa = 2 \times 10^{-4}$. 
The fit with fixed $\sigma =0.5$ gives $U_{\rm c} = 0.36$ with error 
$\kappa = 2\times 10^{-3}$. These $U_{\rm c}$ values are much smaller than 
the one found Ref.\ \onlinecite{bozonos}. Even though it has been emphasized in 
that work that the extrapolated charge gap can be fitted with $U_{\rm c}$ 
varying between $0$ and $2.2$, their best estimate was $U_{\rm c} = 2.2$. The
curve using their parameter values, $a = 45.050$, $c= 6.567$, and  $\sigma = 0.5$ 
is also plotted in Fig.\ \ref{fig:su3_charge_third}. It is clearly seen that this curve is 
well below our extrapolated data even for large $U$.  
The authors of Ref.~[\onlinecite{bozonos}] supported their finding by a theoretical estimate
for the critical coupling $U_{\rm c}$ which they have found to be the order of unity.
Our best estimate at least an order of magnitude smaller. 

\begin{table}[htb]
\begin{tabular}{@{\hspace{1mm}}lc@{\hspace{4mm}}c@{\hspace{10mm}}
     lc@{\hspace{4mm}}c@{\hspace{1mm}}}   \hline \hline
\multicolumn{1}{c}{$U$} &  & $\Delta_{\rm{c}}$ & 
    \multicolumn{1}{c}{$U$ } & & $\Delta_{\rm{c}}$ \\  \hline
0.1  &  & 0.0037(4)  & 2    &  & 0.0065(3)    \\ 
0.25 &  & 0.0039(5)  & 2.25 &  & 0.0101(3)    \\       
0.5  &  & 0.0043(4)  & 2.5  &  & 0.0183(2)    \\       
1    &  & 0.0048(4)  & 2.75 &  & 0.0360(2)    \\
1.25 &  & 0.0049(4)  & 3    &  & 0.0702(2)    \\  
1.5  &  & 0.0051(4)  & 3.5  &  & 0.2090(1)    \\  
1.75 &  & 0.0053(3)  & 4    &  & 0.4421(1)    \\                   
\hline \hline 
\end{tabular}
\caption{Extrapolated values of the charge gap in the thermodynamic
limit for the one-third-filled SU(3) Hubbard model.}
\label{table:su3_third}
\end{table}

Since the numerical results on the gap do not give a definite answer whether $U_{\text{c}}$ is finite
or not, while Ref.\ \onlinecite{bozonos} gives analytic arguments in favor of a finite critical
value, we looked for further numerical evidence by studying the entropy functions. Our results 
are shown in Fig.~\ref{fig:entropy_su3_half}. The 
one-site entropy possesses a maximum at $U=0$. It shows no anomaly 
for $U>0$. For small $U$, it can be fitted with the a quadratic
$U$ dependence with $s_0=1.9076$ (the exact value is 
$s_0 = 3\ln 3 - 2\ln2$) and $A = 0.0138$ with $\kappa=2\times 10^{-7}$.

\begin{figure}[htb]
\includegraphics[scale=0.35]{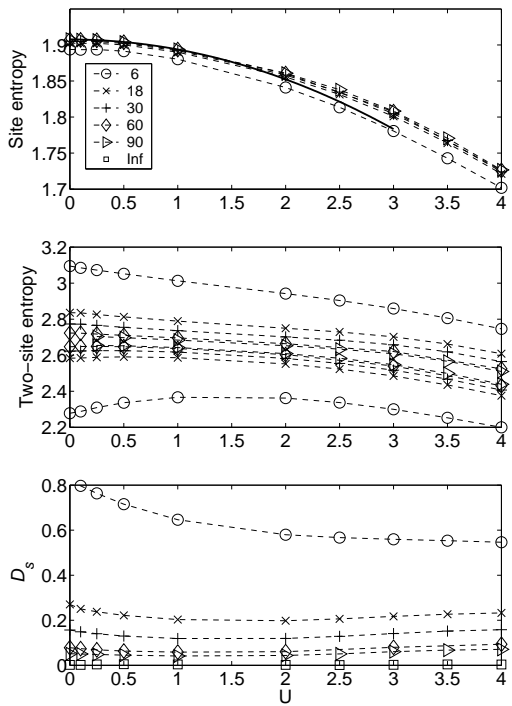}
\caption{Entropy functions plotted as in Fig.\ \ref{fig:entropy_su2}, but 
for the one-third-filled SU(3) Hubbard model.}
\label{fig:entropy_su3_third}
\end{figure}

\begin{figure}[htb]
\includegraphics[scale=0.5]{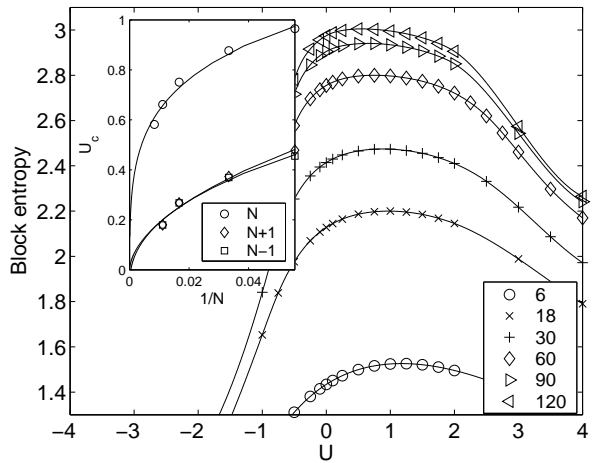}
\caption{Block entropy, $s(l=N/2)$, plotted as a function of $U$ 
for the one-third-filled SU(3) Hubbard model. The solid lines are the result of polynomial fits.
The inset shows the finite-size scaling of $U_c$ determined from the maximum of the block entropy 
for the ground state as well as for excited states with $N+1$ or $N-1$ particles.} 
\label{fig:block_entropy_su3_third}
\end{figure}

A somewhat different behavior is obtained when the entropy of bigger blocks are considered.
This effect is most pronounced when the entropy for a block of length $l=N/2$ is considered.
The block entropy as a function of the block length oscillates now with a period of three, 
the Fourier spectrum has a peak at $q=2\pi/3$, however, this component vanishes in
the large $N$ limit, indicating that the system remains uniform. When the $U$
dependence of the blocks of length $l=N/2$ is taken, this quantity, in contrast to what 
has been seen for the single-site entropy, does not have its maximum at
$U=0$, but at a somewhat larger value. This is seen in Fig. \ref{fig:block_entropy_su3_third}.

The accuracy, the allowed quantum information loss was $\chi=10^{-4}$ in this
calculation, which allowed us to consider chains with $N=120$ sites, but this does 
not influence the shape of the curves. The location of the maximum is shifted by 
less then one percent. The inset in the figure shows the finite-size 
scaling of $U_{\text{c}}$, the location of the maximum of the curves. Similar calculation
have been performed for the excited state with $N+1$ or $N-1$ particles. 
The $U_{\text{c}}$ values obtained from the maximum are also shown in the inset.
Extrapolation to the $N \rightarrow \infty$ limit gives a critical $U_{\text{c}}$
that is smaller than $0.1$. 
Similar result is observed for the two-site entropy. The location of the maximum shifts
to $U=0$ and the dimerization $D_s$ vanishes for all positive $U$.

We have also checked the value of the central charge, which is related to the number
of soft modes, using the initial slope of $s(l)$. While for $U=0$ a value close to 3
is found, our calculation indicates that it is only 2 for $U > 0.5$. For smaller $U$
values much longer chains would be needed to get reliable results.

Therefore, on the basis of our numerical results we conclude that $U_{\rm c}$ is much 
smaller than predicted, and it could even be zero. The ground state is uniform 
as is the case for the half-filled SU(2) Hubbard model.

\subsection{The SU(n) models with $n>3$}

As mentioned in Sec.~III, if an accuracy $\chi=10^{-5}$ is required, relatively short 
chains only could be studied for $n>3$. For the SU(4) model we calculated the energy 
gaps for large $U$ values only to compare our results with those obtained previously by 
quantum Monte Carlo calculation.\cite{bozonos} Here again---similarly to what has 
been seen for $n=3$---the values obtained for the gap are somewhat larger than those 
reported in Ref.\ \onlinecite{bozonos}. For $U=4$ we find $\Delta_{\rm s} = 0.05(3)$
and $\Delta_{\rm c} = 0.21(2)$ for the extrapolated gaps. Since this small but finite 
$\Delta_{\rm s}$ does not allow to draw the firm conclusion that the spin gap vanishes, 
the entropy analysis has been performed. 
 
The entropy functions for the quarter-filled SU(4) model show similar qualitative 
behavior as the half-filled SU(2) or the one-third-filled SU(3) models. The one-site entropy 
possesses a maximum at $U=0$ and for small $U$ it can be fitted with a quadratic form 
with $s_0=2.2353$ (the exact value is $s_0 = 4\ln 4 - 3\ln3$) and 
$A= 0.138$ with error $\kappa=3\times10^{-4}$. None of the entropy functions show 
any anomaly for $U>0$. The dimerization of the two-site entropy seems to
scale to zero, since for any $U$ it scales to a value less than $D_s = 0.02(3)$. 
Although for finite systems the Fourier spectrum of the block entropy $\tilde s(q)$ 
has a peak at $q=\pi/2$, it scales to zero in the thermodynamic limit. The ground 
state is in fact uniform.

Due to the similarity in the behavior of the half-filled SU(2), one-third-filled SU(3), and 
quarter-filled SU(4) models, we suggest that---in disagreement with the result obtained 
by Assaraf {\sl et al}.\cite{bozonos} who predicted vanishing charge gap for 
$U<2.8$---the charge gap opens exponentially slowly at $U_{\rm c}=0$.

The entropy functions of the $1/5$-filled SU(5) model have been calculated for chains up to 
$N=30$ lattice sites only. Although no reliable finite-size scaling could be performed,
our results indicate similar features as seen in the other $1/n$-filled models.
The one-site entropy possesses a maximum at $U=0$ and it is a smooth function of $U$.
We expect, therefore, that in this case as well the Mott transition occurs at
$U_{\rm c}=0$.

\section{Conclusion}

In summary, we have studied the one-dimensional SU$(n)$ ($n=2,3,4$, and 5)
Hubbard model using the density-matrix renormalization-group method.  
Excitation gaps, $l$-site entropies, and Fourier spectrum of the block entropy
have been calculated numerically.
The numerical accuracy has been controlled by the quantum information loss
$\chi$ which also sets a limit on the largest chain lengths whose properties
could be determined with an \emph{a priori} defined accuracy. Our method has 
been tested on the SU(2) Hubbard model in order to see similarities 
and differences when the $n>2$ cases are studied.

First, the model at half filling has been considered. The prediction\cite{marston} 
based on a large-$n$ analytical calculation, that the SU($n$) chain is bond dimerized
for even $n > 2$ has been numerically verified, and it has been shown that exactly
the same behavior is found for odd $n$. For the half-filled models both the spin and
charge excitation gaps are finite for all finite Coulomb repulsion $U$, and
they open exponentially slowly, indicating a Kosterlitz-Thouless transition
at $U_{\rm c}=0$.  Since finite-size scaling is notoriously difficult in the neighborhood 
of a Kosterlitz-Thouless transition, we have studied the behavior of several entropy
functions. They show no anomalies for $U>0$ which also supports that $U_{\rm c}=0$. 
This is, however, not a usual metal-insulator transition. The dimerization for $n>2$
has been corroborated by the peak at $q=\pi$ in the Fourier spectrum of the 
length dependence of the block entropy.

Next, $1/n$-filled systems have been studied. It has been shown that
the one-third-filled SU(3) model and the quarter-filled SU(4) model behave exactly 
in the same way as the half-filled SU(2) model. Although the length dependence of 
the block entropy shows a characteristic oscillation with a period of $1/n$, the
corresponding Fourier components vanish in the thermodynamic limit, confirming that
the ground state is spatially uniform. The spin gap vanishes while the charge gap 
opens exponentially slowly for $U>0$. Moreover, the best fit gives a value for 
$U_{\rm c}$ that is very close to zero. The location of the transition has been checked
by studying the $U$ dependence of various entropy functions. While the one-site 
entropy has its maximum at $U=0$, this is not the case for larger blocks. 
Nevertheless, their maximum shifts to $U=0$ in the large $N$ limit.
Within our numerical accuracy the opening of the charge gap happens with a 
Kosterlitz-Thouless type transition at $U_{\rm c}=0$. The fifth-filled SU(5) 
Hubbard model shows similar behavior. Of course, a finite but very small 
$U_{\rm c} < 0.1$ cannot be excluded, but even such a small value indicates that
the role of higher order um\-klapp processes needs to be reexamined.

In a recent work,\cite{franca} it has been shown that the entropy profile of the SU(2) 
Hubbard model for different $U$ values as a function of the band filling indicates clearly 
the Mott transition at half filling for any positive $U$. Extension of this work to SU($n$) 
models might provide further evidence or discredit our result that in all cases
studied $U_{\rm c}= 0$. 

\acknowledgments

This research was supported in part by the Hungarian Research Fund (OTKA)
Grants No.\ T 043330, F 046356 and NF 61726 and the J\'anos Bolyai Research Fund.  The
authors acknowledge computational support from Dynaflex Ltd under Grant
No. IgB-32.

\end{document}